\documentclass{ws-p8-50x6-00}

\begin{document}

\title{The CMB - Contemporary Measurements and Cosmology}

\author{A. D. Miller\footnote{H\lowercase{ubble} F\lowercase{ellow}}}

\address{Department of Astronomy and Astrophysics, The University of
Chicago, 5640 S. Ellis Ave., Chicago, IL 60637, USA\\ 
E-mail: amber@oddjob.uchicago.edu}


\maketitle

\abstracts{Since the discovery of the Cosmic Microwave Background (CMB) in 1965, 
characterization of the CMB anisotropy angular power spectrum has
become somewhat of a holy grail for experimental cosmology. Because CMB 
anisotropy measurements are difficult, the full potential 
of the CMB is only now being realized. Improvements in
experimental techniques and detector technology have yielded an explosion of progress
in the past couple of years resulting in the ability to use measurements of 
the CMB to place meaningful constraints on cosmological parameters. In this review,
I discuss the theory behind the CMB but focus primarily on
the experiments, reviewing briefly the history of CMB anisotropy measurements and
focusing on the recent experiments that have revolutionized this
field. Results from these modern experiments are reviewed and the
cosmological implications discussed. I conclude with brief
comments about the future of CMB physics.}

\section {Introduction}

The Cosmic Microwave Background (CMB) consists of photons emitted in the early
hot, dense phase of the Universe before the formation of structure by
gravitational collapse. CMB photons are the oldest remaining
photons in the universe; in effect a snapshot of the initial
conditions for structure formation. The CMB is therefore a valuable
tool for connecting the structure we see in the universe today with
theories predicting its origin. 

At times earlier than $z\approx 1000$, the universe consisted of a
hot dense plasma in which electrons were electromagnetically coupled
to baryons, photons were coupled to electrons via Compton scattering,
and the universe was opaque to radiation. When the
universe cooled enough to form the first neutral
hydrogen atoms, the number density of free-electron scatterers dropped
precipitously and the photons effectively decoupled from the baryons.
Small inhomogeneities in the initial density field, which existed before decoupling,
seeded the gravitational collapse of baryonic mater, resulting in the formation of
all observable structures in the current universe.
The photons were left to free stream and cool, interacting very little for
the remaining history of the universe. These CMB photons therefore
provide the cleanest known probe of the early universe;
anisotropies in the observed CMB temperature field directly correspond to the tiny
fluctuations in the initial density field.

\section{Anisotropy in the CMB - Theoretical Overview}

The CMB is remarkably uniform, but small inhomogeneities in the density field of the early
universe are revealed by temperature fluctuations of order one part in $10^5$ in the
currently observable CMB. Information from measurements of these fluctuations is quantified by constructing the angular power spectrum of the fluctuations discussed below. Since different models of structure formation predict different power spectrum characteristics, observations can be used to test cosmological predictions. 

\subsection{The Angular Power Spectrum}

CMB temperature fluctuations
on the sky can be
described by an expansion in spherical harmonics, 
\begin{equation}
{T(\theta,\phi)}=\sum_{l,m}a_{lm}Y_{lm}(\theta,\phi). 
\label{cmb_temp}
\end{equation}
For Gaussian fluctuations, the multipole moments of the temperature field
are fully described by their angular power spectrum, 
\begin{equation}
\big<a_{lm}^*a_{l\prime m\prime}\big>=\delta_{ll\prime}\delta_{mm\prime}C_l
\label{random_phase}
\end{equation}
The angular power spectrum can be related to the
autocorrelation function, $C(\theta)$, as
\begin{equation}
C(\theta)={1\over 4\pi}\sum_l(2l+1)C_lP_l(\cos\theta)
\label{autocorrelation2}
\end{equation}
where the $P_l$'s are the Legendre Polynomials. Since the Legendre
Polynomials are orthogonal, the individual values of $C_l$ can be
recovered by multiplying $C(\theta)$ by a Legendre Polynomial and
integrating over $\cos\theta$.

It is customary to plot the angular power spectrum, $C_l$, in terms of $(\delta
T_l)^2$, the power per logarithmic interval in $l$, 
\begin{equation}
(\delta T_l)^2={1\over 2\pi}l(l+1)C_l
\label{delta_t_squared}
\end{equation}
or as the square root of this quantity, 
\begin{equation}
\delta T_l=\sqrt{l(l+1)C_l\over2\pi}.
\end{equation}

\begin{figure}[t]
\epsfxsize=30pc 
\epsfbox{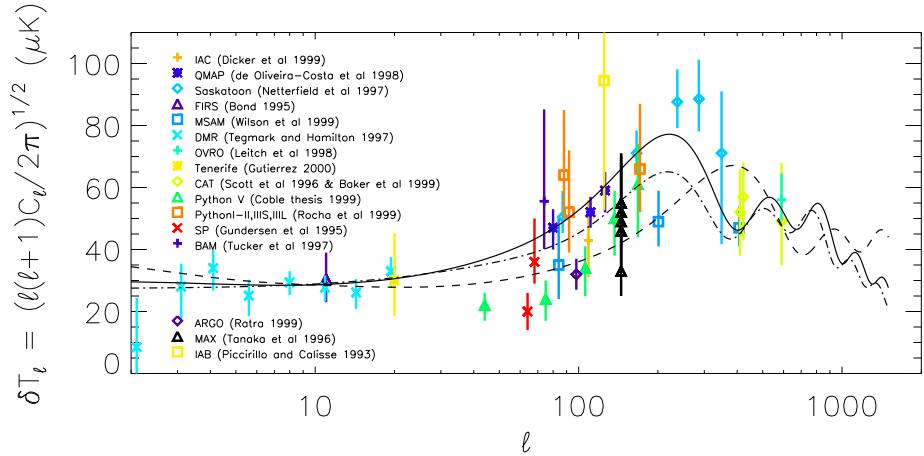} 
\caption{Models for the CMB angular spectrum computed with CMBfast\protect\cite{S_Z}
plotted with CMB data published as of 1999.
The solid line is a $\Lambda$CDM model with
$\Omega_m=0.33$, $\Omega_b=0.041$, $\Omega_{\Lambda}=0.67$, and
$h=0.65$. The dotted line is a standard CDM model with $\Omega_m=1$,
$\Omega_b=0.05$, and $h=0.65$. The dashed line is an open model with
$\Omega_m=0.3$, $\Omega_b=0.05$, and $h=0.65$. Plotted CMB bandpowers 
do not include calibration errors, which are typically at least $10-20\%$.
\label{fig:figure1}}
\end{figure}

\subsection{Primary Anisotropies}

Three representative models for the CMB angular spectrum are shown in 
Figure~\ref{fig:figure1}. 
A typical
model is characterized by a flat plateau at large angular scales rising with $l$ to a first prominent
peak, and followed by peaks of decreasing amplitude at smaller angular scales. The 
series of peaks and valleys is referred to as ``Doppler'' or
``acoustic'' peaks. The primary anisotropies discussed in this section
are those occurring on the
surface of last scattering. We distinguish them from secondary
anisotropies discussed in the next section, which are due to photons
scattering along the line of sight between the surface of last
scattering and the observer. The three fundamental effects that produce
primary anisotropies can be summarized by the equation~\cite{Peacock,Tegmark}, 
\begin{equation}
{\delta T\over T}({\bf r})=\phi({\bf r})+{1\over 3}\delta({\bf r})-{\delta {\bf
v}\cdot  \hat{\bf r}}.
\label{primary_anisotropies}
\end{equation}

The first term represents a redshift due to the gravitational potential,
$\phi$. Photons in overdense regions must climb out
of potential wells in order to reach an observer. Dense regions,
therefore appear cooler by an amount, $\delta T/T=\phi({\bf r})$. 

The second term is the adiabatic contribution. 
The physical mechanism behind this term 
is presented by Hu, Sugiyama, and Silk~\cite{H_S_S} and is explained in
detail there. The basic idea is the following. Before decoupling, the
photons and baryons are tightly coupled (due to Compton scattering
between electrons and photons and Coulomb interactions between electrons
and baryons) and can be treated as a single fluid.
In the context of a cold dark matter (CDM) model,
Gaussian random fluctuations in the
dominant CDM component create potential wells into which the
photon-baryon fluid falls. The potential wells compress the fluid,
the photon temperature increases, and the radiation pressure rises enough to
overcome the compression and to cause a rarefaction. These oscillations
proceed until recombination when the photon-baryon coupling breaks
down. The denser regions on the last scattering surface will therefore
be hotter while the less dense regions are cooler. Following inflation
but before recombination, the horizon size increases and oscillation modes of
increasing size enter the horizon. These modes are
frozen in at last scattering, leaving peaks in the angular spectrum
corresponding to modes caught at extrema at that epoch. The smallest
wavelength modes have had the most time to evolve before being frozen.
Odd peaks represent the compression maxima while even ones result from
rarefaction maxima. The first ``Doppler Peak'' represents
the mode that has just reached its first compression maximum at
the time of last scattering. The density
perturbations can be parameterized in terms of the density contrast,
\begin{equation}
\delta={\delta\rho\over\rho},
\label{overdensity}
\end{equation}
where $\rho$ is the average density and $\delta\rho$ is the local density
perturbation. Assuming that the
universe was matter-dominated at the epoch of decoupling, the density
$\rho\propto T^3$ so
\begin{equation}
{\delta T\over T}_{\rm adiabatic}={1\over 3}\delta.
\label{adiabatic}
\end{equation}
The longest wavelength modes ($>$ a few degrees) will not have had time
to evolve
significantly by the time of decoupling and will yield information about the
primordial potential fluctuations. Assuming adiabatic initial
conditions, $\delta\approx -2\phi$~\cite{Wayne_thesis}, or 
\begin{equation}
{\delta T\over T}_{\rm adiabatic}=-{2\over 3}\phi
\label{sachs-wolfe1}
\end{equation}
In this case, the first and second terms partially cancel yielding 
\begin{equation}
{\delta T\over T}={\phi\over 3}.
\label{sachs-wolfe}
\end{equation}
This is called the Sachs-Wolfe effect and is responsible for the
plateau in the
angular spectrum seen in Figure~\ref{fig:figure1} at
large angular scales (low {\it l}).

The third term represents the 
Doppler contribution, which is due to the combination of the Earth's motion
with respect to the CMB and the motion of the matter on the surface of
last scattering. In a universe in which the baryon contribution is large
enough to be consistent with nucleosynthesis, the velocity contribution
is subdominant~\cite{Wayne_thesis}.

\subsection{Secondary Anisotropies}

In addition to the primary anisotropies that imprint temperature
fluctuations on the CMB before decoupling, several processes also exist
after last scattering that imprint detectable secondary
anisotropies. These can be broadly divided into two categories; those
due to gravitational effects (early and late Integrated Sachs-Wolfe (ISW) 
effects and the Rees-Sciama effect), and those due to reionization at later
times in the history of the universe. The mechanisms governing secondary 
CMB anisotropies are discussed elsewhere~\cite{Wayne_thesis,Tegmark} but 
some observational consequences of
these effects are presented in the next section.

\subsection{Observational Signatures}

Measurements of the CMB angular spectrum are a powerful probe of the
underlying cosmology because the detailed shape of the power spectrum
depends on the specifics of the cosmological model. In particular,
parameters such as the energy density of the universe relative to
the critical value, $\Omega$,
the fraction of that density in baryons, $\Omega_b$, matter,
$\Omega_m$, curvature, $\Omega_k$ and due to a cosmological constant,
$\Omega_{\Lambda}$, as
well as the Hubble constant, $H_0$, the tilt, $n$, and the reionization
fraction, $\tau$ each leave an observable signature on the CMB angular
power spectrum. Discussed below are some of the
cosmological parameters which, according to the most popular class of
adiabatic models, can be estimated by careful measurement of the angular
power spectrum. 

$\bullet$ {\bf $\Omega_k$:}
The location of the acoustic peaks in $l-$space depends most sensitively
on the curvature of the universe through a simple
projection effect~\cite{JKKS}. The scale of the sound horizon in the photon-baryon
fluid before decoupling sets the maximum size of a causally connected
region, corresponding to the angular size of the first acoustic
peak. The
radiation density is fixed by the FIRAS measurements of ${\rm T}_{\rm
CMB}$ today~\cite{FixsenCOBE} so the physical size of the sound
horizon depends primarily on $\Omega_mh^2$ and can be used as a standard
ruler. The measured angular scale at which this peak is detected
therefore sheds light on the background geometry. The measured angular
scale for an object of fixed size will be larger for a universe with
positive curvature than for a flat universe and will be smaller for a
universe with negative curvature. When detailed calculations
are performed, the first acoustic peak is shown to be located at
$l\approx 220$ for a large
class of flat models and $l\approx 350-400$ for a large class of open
models.

$\bullet$ {\bf $\Omega_m$:} 
The primary signature on the CMB of the matter density is a shift in the
position of the peak due to the curvature effect discussed above. A
lower value of $\Omega_m$ in the absence of a compensating higher value
of $\Omega_{\Lambda}$ will shift the peak to higher $l$ values. In the
context of a flat universe, a lower value of $\Omega_m$ implies a larger
$\Omega_{\Lambda}$ or $\Omega_{\it DE}$\footnote{$\Omega_{DE}$ refers to a 
component of dark energy with negative pressure but in which the equation of state,
$w>-1$ in contrast for that of $\Lambda$ for which 
$w=-1$~\cite{Caldwell,Huey}}.

$\bullet$ {\bf $\Omega_{\Lambda}$ or $\Omega_{DE}$:}
If the universe is dominated by a negative pressure component such as 
a cosmological constant or some other form of dark energy,
the lowest multipoles will be enhanced by the late ISW 
effect. The presence
of a negative pressure component provides the necessary energy density
to accommodate
a flat universe in which $\Omega_m\approx 0.3$, consistent
with other measurements~\cite{Freedman}. In a flat universe,
the early ISW effect leads to an increase in amplitude of the first peak
with increasing cosmological constant. This dependence is due to the
fact that the baryon-to-photon ratio before decoupling is lower in models
that are dominated at the present epoch by some form of dark energy
than it is in flat models in which the
energy density is due entirely to matter. This is
the same effect as described above for a lower value of $\Omega_m$ but
peak amplitude is boosted without the accompanying shift to
smaller angular scales characteristic of open models. 

$\bullet$ {\bf $\Omega_b$:}
Under the assumption that the universe is dominated by dark matter
at early times, the photon-baryon fluid oscillates under the driving
force of gravity and the restoring force of photon pressure. The
effective mass term of the photon-baryon fluid depends on the
baryon-to-photon ratio. As the baryon contribution is enhanced, the
additional mass provided by the baryons increases the compression of the
fluid in the well and enlarges the compression peaks (odd peaks). In
addition, the photons diffuse in the photon-baryon fluid by an amount
determined by the ionization history and the baryon
content~\cite{H_S_S}. The effect of this diffusion is a damping of structure on
small scales as photons from over dense regions mix with those from
under dense regions. The extent to which the angular spectrum is
suppressed at higher multipoles will, therefore depend on the baryon
content. 

$\bullet$ {\bf $H_0$:}
During radiation domination, radiation pressure induces a
decay of  the gravitational potential, boosting the height of the
acoustic peaks. If $H_0$ is relatively lower, matter-radiation equality
happens later and the peaks are enhanced. In
other words, one interpretation of a high first acoustic peak is a
relatively lower value of the Hubble constant which implies a delay in
matter-radiation equality. The
location of the peaks also depends weakly on the Hubble constant. The
value of the Hubble constant affects the distance to the last
scattering surface. A larger distance to the last
scattering surface will result in features of known physical extent
appearing at smaller angular scales, which translates to a shift in the
peaks to higher $l$. 

The currently favored class of models are
inflation-inspired adiabatic models in which potential wells influence
all particle species in the same way;
the fractional fluctuation in the number density of each species is
independent of species. Most of the above discussion assumes models of
this type. Isocurvature models, on the other hand, arise from initial
conditions where there are no
potential wells. Fluctuations in the photon density balance fluctuations
in the other particle species in order to maintain zero curvature in the
initial state. As in the adiabatic case, oscillations are produced in
the photon-baryon fluid but they are $90^{\rm o}$ out of phase with respect
to those expected from adiabatic initial conditions. 
The first acoustic peak is therefore 
expected to be found at $l\approx 350$ $\Omega^{-1/2}$ rather than
$l\approx 200$ $\Omega^{-1/2}$ as is expected for adiabatic 
models~\cite{H_S_S}. Isocurvature conditions can arise
from cosmological defects
(strings, monopoles, and textures)~\cite{Magueijo,Crittenden}. An
essential signature of most
defect models is that temporal phase coherence for Fourier modes of a
given wavenumber, which is responsible for driving the acoustic peaks,
is destroyed by the non-linear evolution of the defect network. The
result is that the fluctuation power is smeared out and the series of
acoustic peaks is converted into a single broad peak, wider than the
observed first Doppler peak and shifted to higher $l$. Simple
isocurvature models are not consistent with contemporary CMB measurements.

\section{CMB Anisotropy Measurements - Experimental Challenges}

Many experiments have been designed to measure
the anisotropy in the CMB. These measurements are difficult
due to the small relative amplitude of the effect. Fluctuations exist at
the level of a few parts in $10^5$ of a ${\approx 3}$ K signal that is already
difficult to distinguish from atmospheric noise, emission from Galactic and
extragalactic sources, and systematic errors inherent to the
experiments. A successful CMB experiment must have high sensitivity
and low and well understood systematic errors.

There are three general approaches that have been taken in order to measure
anisotropy in the CMB. The first is to make a map of the fluctuations and to
extract the power spectrum from the map. Examples of map-making experiments
include satellite experiments such as COBE ~\cite{Smoot}, NASA's recently launched MAP,
and ESA's upcoming Planck. Mapping has also been a popular technique for
balloon-borne experiments such as QMAP~\cite{Devlin98}, BOOMERanG~\cite{deBernardis_inst},
and MAXIMA~\cite{Lee}. CMB maps are characterized by complex scan patterns,
facilitating the comparison of each pixel (containing a temperature measurement and an
associated uncertainty) with each neighboring pixel. Mapping instruments are generally
made up of a single
collecting telescope dish as a primary mirror with secondary and sometimes
tertiary mirrors used to focus the radiation into the receiver. Receivers
are single or multiple element radiometers generally constructed using 
HEMT~\cite{Pospieszalski} detectors at frequencies below ~90 GHz and
bolometers at higher frequencies.

The second class of CMB experiments are those making use of a scanning
or beam switching technique to compare signals at difference locations
on the sky. These experiments are constructed in much the same way as are map-making
instruments (also generally single-dish telescopes using HEMTs and bolometers) but
the scan strategy is optimized for probing the power
spectrum directly without the intermediate step of making a map. This is an approach 
often taken by ground-based experiments, simplifying the scan to a constant 
elevation in order to keep the atmospheric temperature as uniform as possible over the  
scan. Examples of this type
of experiment are Saskatoon~\cite{Wollack93},MSAM~\cite{Fixsen1996}, MAX~\cite{tanaka}
Tenerife~\cite{Romeo2000}, PYTHON~\cite{Coble1999}, VIPER~\cite{Peterson1999}, and
MAT ~\cite{Miller,Torbet}.

The third general class of CMB experiments are those using interferometry. Since the natural
data products of an interferometer are components of the Fourier transform of the 
sky brightness distribution, interferometry is a natural way to measure the angular
power spectrum of the CMB. Interferometers also
have the advantage of low systematic errors, easy removal of point sources, and good atmospheric
rejection. Interferometers require a phase-preserving detector so generally HEMT amplifiers or
SIS mixers are used rather than the more sensitive bolometers. Examples of 
interferometric CMB experiments are CAT~\cite{Baker1999}, DASI~\cite{Leitch}, 
IAC\cite{Harrison2000}, and CBI~\cite{Padin2001}.

\section{Early CMB Experiments}

The Hot Big Bang model was introduced by Gamow in 1946~\cite{Gamow}
and the existence of a cosmic radiation background was first predicted by
Alpher and Herman in 1948~\cite{Alpher}. These ideas were largely
ignored, however until the sixties when Dicke and Peebles
worked out a new hot big bang model and predicted the existence of a
resulting thermal background relic. Roll and Wilkinson
were preparing an experiment to search for this radiation when it was
detected by Penzias  and Wilson in 1964  as excess
noise in their communications antenna. The discovery~\cite{PandW} and the
explanation~\cite{DPRW65} were published in companion papers in 1965
and the field of CMB physics was officially on the map.
The CMB spectrum was definitively measured over the frequency range
between 70 and 630 GHz by the Cosmic
Background Explorer (COBE) Satellite's Far Infrared Absolute
Spectrophotometer (FIRAS) instrument and found to be that
of a thermal blackbody with a temperature of $2.728\pm 0.004$  
K at the $95\%$ confidence level~\cite{FixsenCOBE}.

While the measurement of the CMB and the identification of the radiation
as such a strikingly good black body was a triumph for the big bang model,
information about the formation history of the large scale structure in
the universe is contained not in the absolute measurement of the CMB but
in measurements of the anisotropy. Had no anisotropy been detected, the
entire model of structure formation growth by the collapse of density
inhomogeneities would have to have been reformulated. The COBE DMR detection
of these anisotropies on angular scales of several degrees~\cite{Smoot}
therefore marked another large milestone in the field and encouraged
other experimentalists to look further.

COBE measured fluctuations on scales that reflect the primordial fluctuation
field; those fluctuation modes that had not yet had time to complete
an oscillation in the potential well before crossing the horizon. These
measurements therefore normalize the CMB spectrum and can be used to connect
with measurements of large scale structure.

Since the location in {\it l}-space of the first peak contains the most direct
known measurement of the geometry of the universe and the location and height
of this peak contain a wealth of information about other cosmological parameters,
the next challenge was to determine the shape of the angular power spectrum
on scales near the location of the predicted first peak. Since the COBE DMR detection,
measurements of the CMB have largely focused on this goal. While 
descriptions of individual experiments will be confined to recently published
results, it is important to note the immense amount of progress that was
made by early experiments.
Figure~\ref{fig:figure1} shows CMB data published before 1999.
It is clear that there is evidence for a rise in the spectrum up to
$l \approx 200$ and a fall above $l\approx 400$ but no single instrument
had detected and localized the peak. Since each instrument has a calibration 
error on the order of $10\%$
or more and the error bars are large, measurements at different angular scales
made with different instruments are difficult to compare in a quantitative way.
It was therefore important that single instruments (with the same systematic
errors and the same calibration uncertainties) measured the angular spectrum 
over the entire range of multipoles spanning the predicted location of the peak.

\section {Modern CMB Experiments}

The dream of the CMB being
used to place meaningful constraints on cosmological parameters has come true over the
past couple of years. The location
of the first Doppler peak has been conclusively measured, the second and arguably
third peaks have been detected, and constraints on parameters have been inferred
from the location and relative heights of these peaks. It is therefore useful
to discuss in more detail these measurements and the instruments that have made
them possible. Data from modern CMB experiments is plotted in Figures 2a-2c and 
Table 1 shows experimental parameters and the best fit cosmological models for 
each of the experiments.

\begin{figure}[t]
\epsfxsize=14pc 
\epsfbox{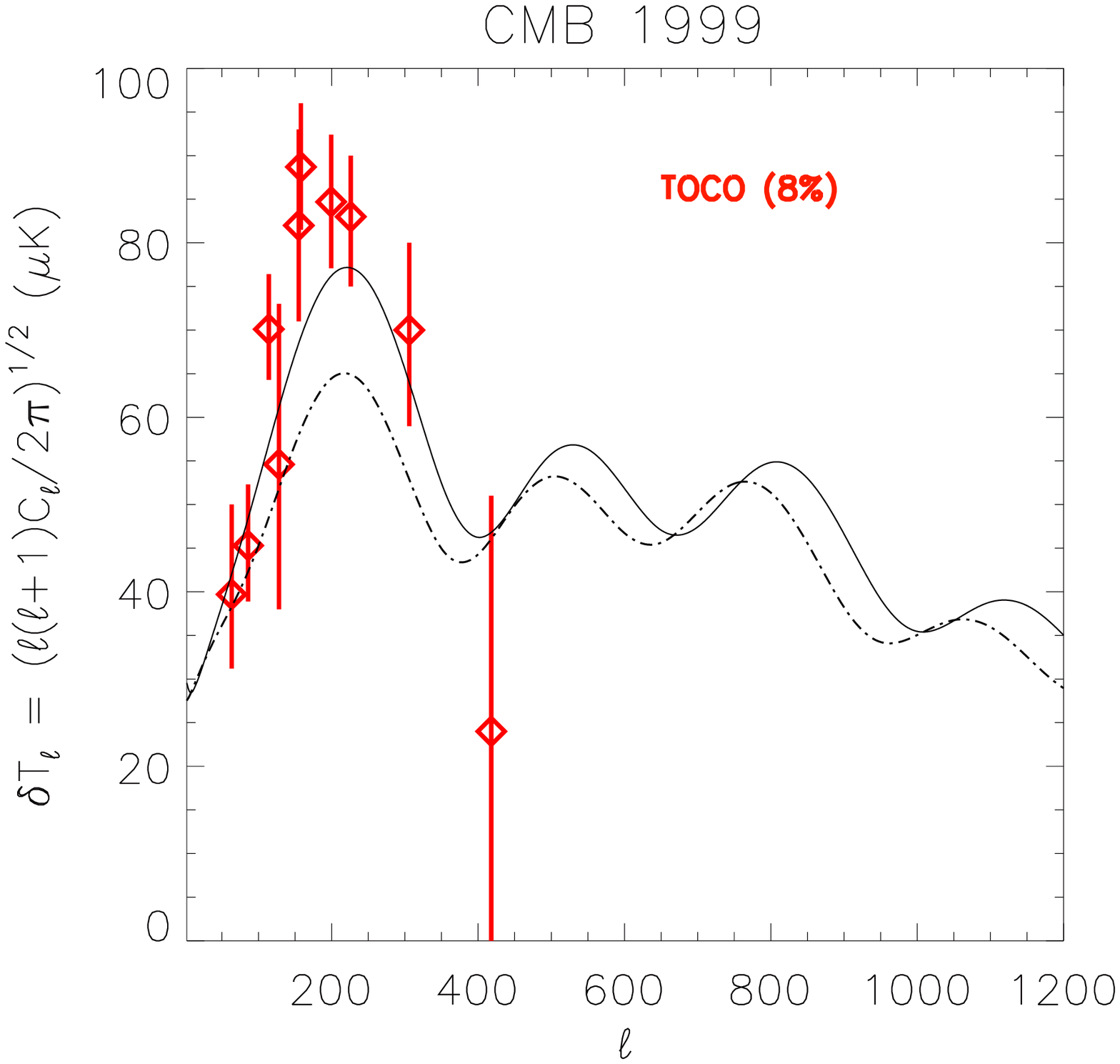} 
\epsfxsize=14pc
\epsfbox{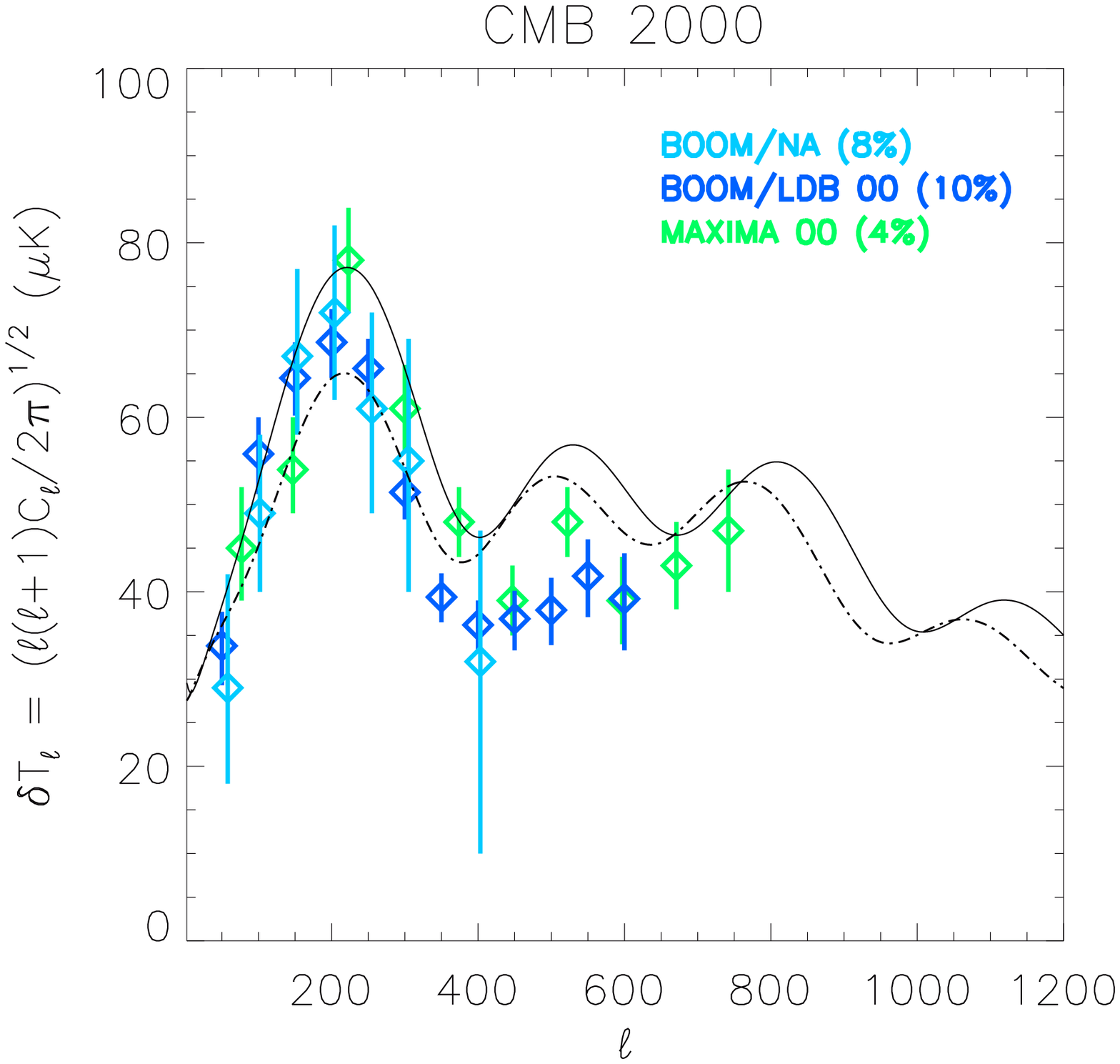}
\epsfxsize=14pc
\epsfbox{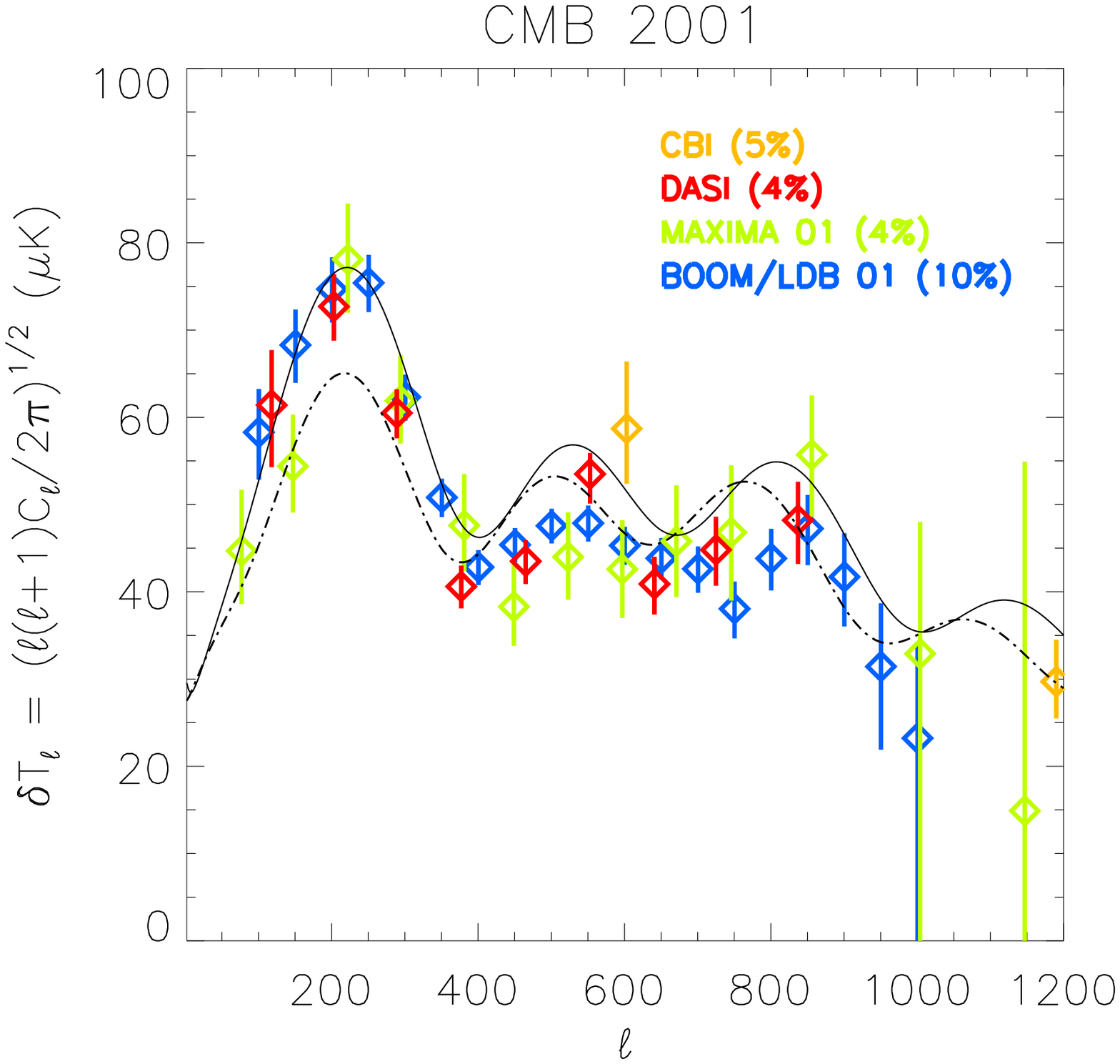}
\epsfxsize=17pc
\epsfbox{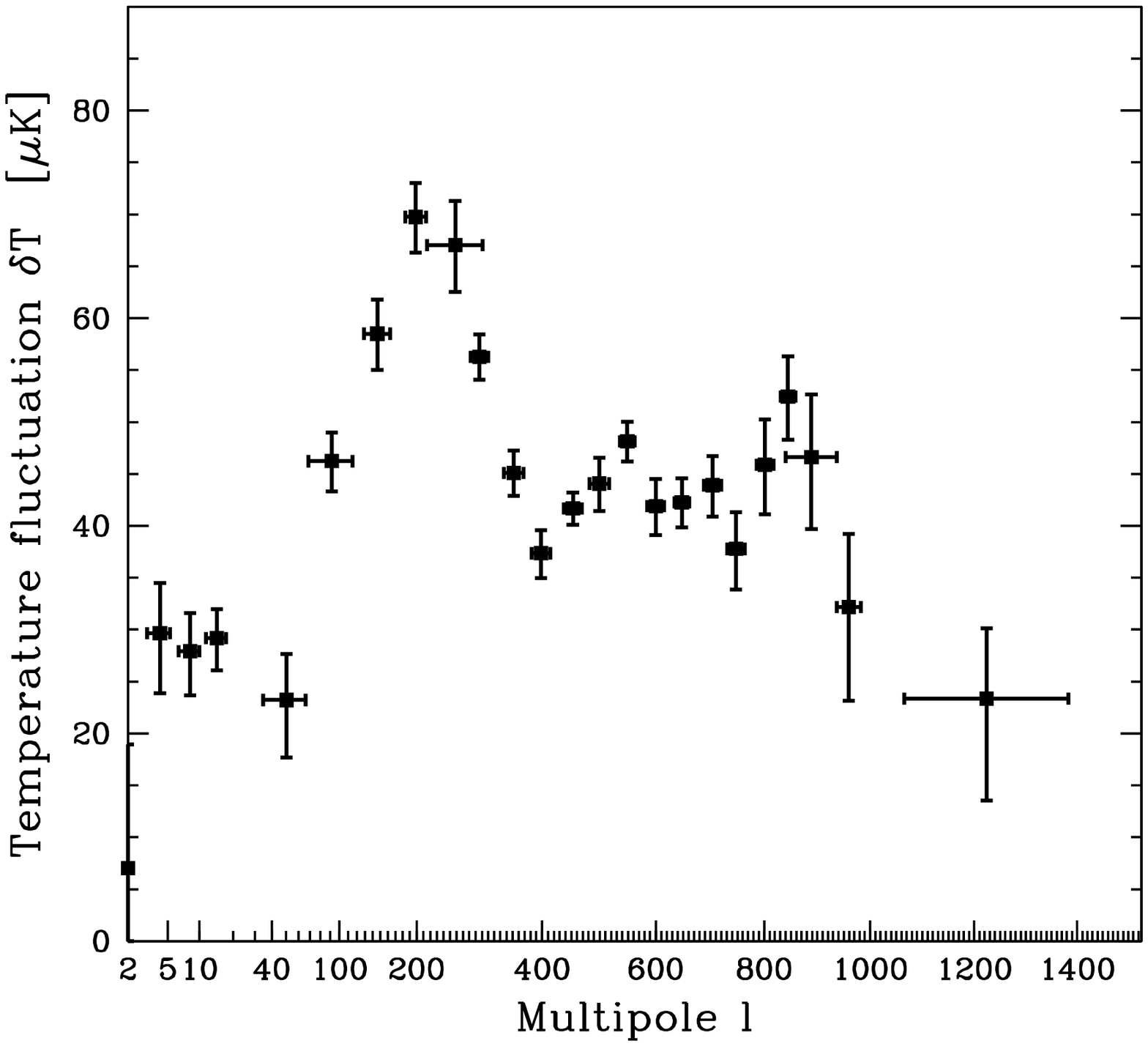}
\caption{{\bf a} (upper left): The MAT/TOCO data\protect\cite{Miller,Torbet}
{\bf b} (upper right):Data from BOOM/NA\protect\cite{Mauskopf}, 
BOOM/LDB\protect\cite{deBernardis}, and MAXIMA\protect\cite{Hanany} published in 2000.
{\bf c} (lower left): DASI\protect\cite{Halverson}, CBI\protect\cite{Padin2001a}, 
the BOOM/LDB 2001 analysis\protect\cite{Netterfield_Boom}, and the MAXIMA 
2001 analysis\protect\cite{Lee2001}. The BOOM/LDB 2001 analysis differs from the 2000 
analysis in that
the number of channels analyzed were increased from one to four,
the data from the second half of the flight was added, the effective beam 
size was adjusted from
$10.0\pm0.1\prime$ to $12.9\pm1.4\prime$, and the overall calibration
was raised by $10\%$ in $C_l$\protect\cite{Netterfield_Boom}. The MAXIMA 2001 analysis
 differs from the 2000 analysis in that, the number of channels 
used was dropped
from four to three (the 240 GHz channel was removed leaving three channels at 150 GHz).
Also, the 2000 analysis made use of the entire map binned into pixels of size
$5\prime \times5\prime$ while in 2001 analysis was restricted to the central portion of the map
and the data was re-binned into pixels of size $3\prime\times3\prime$\protect\cite{Stompor}.
{\bf d} (lower right) CMB combined bandpowers\protect\cite{wang} obtained by combining much
of the data from Figure 1 and all data from Figure 2 with the exception of BOOM/LDB 2000 and 
MAXIMA 2000. Combined bandpowers include calibration and beam uncertainties from individual 
experiments.
Note, models in each panel are the two $\Omega_{TOT}=1$ models from Figure 1 
and numbers in parentheses are calibration uncertainties. 
\label{fig:figure2}}
\end{figure}

The first detection of the first Doppler peak with a single instrument came from 
the Mobile Anisotropy Telescope
on Cerro Toco (MAT/Toco)~\cite{Miller}, a single dish scanning experiment using HEMT~\cite{Pospieszalski} and 
SIS-based~\cite{Kerr} receivers.
MAT operated at the base of Cerro Toco mountain at an altitude of 17,000 ft. (5200 m) in the 
Atacama desert in Northern Chile.
Data published in 1999 from this experiment (see Figure 2a)
localized the peak in the spectrum to $l\approx 200$,
giving support to adiabatic, inflationary models and ruling out the most common
class of isocurvature and texture models. The amplitude of the first peak as measured
by Toco was also suggestively high. As seen in figure~\ref{fig:figure2} the data
does not fit the standard CDM model very well and is a better fit to a $\Lambda$CDM
model.

The next experiment to detect a peak in the CMB power spectrum was the BOOMERanG 
experiment~\cite{deBernardis_inst}.
BOOMERanG is also a single dish experiment with multiple detectors operating at
higher frequency and higher angular resolution than the Toco experiment. It uses
bolometers rather than HEMTs or SISs and operates from a balloon platform. BOOMERanG
was the first CMB experiment to fly on a long duration balloon from Antarctica. Before
flying in Antarctica, the instrument underwent a test flight in North America. Data
from the test flight (BOOM/NA)~\cite{Mauskopf} was published early in 2000 and showed striking
agreement with
the Toco data in the location of the peak (see Figures 2a and b). The amplitude of the peak 
is lower than that measured by Toco but agrees within the uncertainties.
Also published in 2000,
data from the BOOMERanG long duration flight (BOOM/LDB)~\cite{deBernardis} measured, 
for the first time, entire rang of {\it l}-space covering the
predicted locations of the first and second peaks in the angular power spectrum.
This data caused some speculation at the time of publication because while power was clearly
detected in the region of the second peak, the amplitude was lower
than predicted by the cosmological models that were at the time becoming standard.

Also published in 2000, data from the MAXIMA experiment
again confirmed the location of the first peak and measured the region of {\it l}-space
covering the second peak~\cite{Hanany}. MAXIMA~\cite{Lee} is a balloon-borne mapping 
experiment that flew from North America. While MAXIMA saw slightly
more power in the region of the second peak than BOOMERanG did, joint fits to the MAXIMA
and BOOMERanG data yielded a value of $\Omega_b h^2$ higher than predicted from Big Bang
Nucleosynthesis~\cite{TegandZal,Bond2}. Analyses suggested that within the standard cosmology
and standard BBN, this value could not be accommodated~\cite{Burles}, and alternative
theories were proposed such as non-standard reionization histories~\cite{Hannestad}, and 
the possibility of evolution in the electromagnetic entropy since the time of BBN~\cite{Kaplinghat}.

Data published in 2001 resolved this discrepancy. Three data sets were released in the
same weekend, covering the range of {\it l}-space spanning the first three Doppler
peaks, which agreed remarkably well.
Figure 2c shows data from the first year of DASI 
observations along with re-anlyses of the BOOMERanG and MAXIMA 2000 data. Differences 
between the 2000 and 2001 analyses for these experiments are outlined in the caption for
Figure 2.

DASI (Degree Angular Scale Interferometer) is a very different instrument from those
discussed above. Rather than being a single dish mapping or scanning telescope, it
is a compact interferometer designed specifically for the purpose of measuring anisotropy
in the CMB. Like Toco, it uses HEMT amplifiers and operates from the ground (in this 
case the South Pole). DASI was designed to operate as a sister experiment to another interferometer, CBI (Cosmic Background
Interferometer) with larger dishes and larger dish separation than DASI, therefore
designed to probe smaller angular scales (see Figure 2c). CBI operates from the a desert plateau  in the Northern Atacama desert very close to the Toco site. 

Focusing on figure~\ref{fig:figure2}a-c, there are a couple of subtleties
to note. First, the data plotted in figures 2a-2c
do not include calibration uncertainties. These uncertainties are listed in parentheses
in the upper right of each figure. One is free to shift each entire data set up or down 
by the calibration uncertainty. In addition, the BOOMERanG and MAXIMA data sets 
have uncertainties in the effective beam size that are not included
in the plotted error bars. These uncertainties depend on $l$ (grow with increasing $l$) so
the effect is to tilt the spectrum. See Netterfield {\it et al.} for an illustration of 
the amplitude of the tilt corresponding to the 1 sigma uncertainty in the BOOMERanG beam 
of $\pm 13\%$. Beam and pointing errors for the MAXIMA experiment are listed as a function 
of $l$ in Lee {\it et al.} and range from $0\%$ at $l=77$ to $^{+25}_{-18}\%$ and 
$\pm 10.2\%$ at $l=1147$ for beam and pointing respectively. Caveats stated, agreement between the three independent measurements covering a large range of angular scales and the CBI result 
with the smallest error bars at the smallest angular scale is impressive. In addition, it is 
reassuring to note that while BOOMERanG and MAXIMA are similar instruments employing 
similar scan strategies, DASI is a completely different type of instrument producing
indistinguishable results. 

Figure 2d shows CMB data as combined by Wang, Tegmark, and Zaldarriaga~\cite{wang}. 
Data have been compressed from an original 105 bandpowers
to the 24 bandpowers shown in the Figure 2d.  Clearly evident are three peaks in 
the CMB power spectrum, the first of which appears at $l\approx 200$, followed by 
a decrease in power at smaller scales. Cosmological implications are discussed below.

\subsection{Cosmological Parameters and Implications}

As discussed above, much of the focus of CMB experiments following COBE has gone into 
characterizing the first Doppler peak in order to constrain the geometry of the
universe. Using the Toco data and the BOOMERanG/NA data alone, Knox and Page performed 
a model-independent fit to the data solving for the location and height of the 
first Doppler peak~\cite{Knox_Page}, localizing it to be at $175<l<243$ (Toco)
and $151<l<259$ (BOOM/NA) both at the $95\%$ confidence level, as predicted
by adiabatic inflationary models. Models predicting the peak location to be at
$l\approx 350-400$ (isocurvature and texture models) are therefore ruled out at the many sigma
level by each data set independently. Combining this data with recent low 
measurements of the matter 
density~\cite{Freedman} we can infer that there
must be a component of dark energy in the universe. This agrees well with 
supernova measurements~\cite{Perlmutter1998,Riess}, which suggest that the
cosmological constant (or other form of
dark energy) plays a significant role in the dynamics of the universe today.

Beyond the location of the first peak, the detection of the other peaks along with
the characterization of their locations and heights relative to each other and to
the first peak shed light on cosmological parameters.
Table~\ref{tab:table1} shows the values
of several cosmological parameters as measured by each of the CMB experiments 
both independently and when combined into a single set of bandpowers~\cite{wang}.
It is clear that individual determinations are consistent with each other and 
with the general picture of a flat, low $\Omega_m$ universe that is currently 
dominated by some form of dark energy.

\begin{table}[t]
\caption{{\bf Top: Experimental parameters} References at top refer to papers describing the
instruments. {\bf Bottom: Best fit cosmological parameters} 
Column two shows the best fit values for TOCO+DMR assuming a prior of $\Omega_b h^2=0.019\pm0.003$ and $h=0.65\pm0.1$\protect\cite{Dodelson}. No formal errors were calculated. Column three shows 
cosmological parameters extracted from BOOM/LDB 2001 data. Error bars are one sigma and assumed priors
are $0.45<h<0.9$, and age $>10$Gyrs\protect\cite{Netterfield_Boom}. Similar fits were done to the 
BOOM/LDB 2000 data assuming the same priors resulting in $\Omega_{TOT}=1.15^{+0.1}_{-0.09}$, 
$\Omega_bh^2=0.036^{+0.006}_{-0.005}$, $\Omega_{CDM}h^2=0.24^{+0.08}_{-0.09}$, $\Omega_m=0.84\pm0.29$, $\Omega_{\Lambda}<0.83$, and $n_s=1.04^{+0.1}_{-0.09}$\protect\cite{Lange}. Fits to BOOM/NA data 
result in $0.85<\Omega_{TOT}<1.25$ at the $68\%$ confidence level and when combined with COBE 
results in a
best fit model of $\Omega_{CDM}=0.46$, $\Omega_b=0.05$, $\Omega_{\Lambda}=0.5$, $n_s=1$, and $h=70$\protect\cite{Melchiorri}. Column four shows the best fit parameters for the MAXIMA data released 
in 2000 combined with COBE\protect\cite{Balbi}. Error bars are at the $95 \%$ confidence level with
$0.4<h<0.9$, age $>10$Gyr, $\Omega_m>0.1$. 
The values of $\Omega_{\Lambda}$ and $\Omega_m$ were calculated combining MAXIMA data 
with Supernova results. Under the same assumptions, fits to the 
MAXIMA 2001 data combined with COBE result in (at the $95\%$ confidence level) 
$\Omega_bh^2=0.0325\pm0.0125$, $\Omega_{CDM}h^2=0.17^{+0.16}_{-0.07}$, 
$\Omega_{TOT}=0.9^{0.18}_{-0.16}$, and $n_s=0.46\tau_c=(0.99\pm0.14)$ assuming that 
$\tau_c\le0.5$\protect\cite{Stompor}. Column five contains the best fit cosmological parameters
to the DASI data\protect\cite{Pryke}. Errors are $68\%$ confidence and priors are $h>0.45$ and
$0.0<\tau_c\le 0.4$. Constraints on $\Omega_m$, and $\Omega_{\Lambda}$ also assume $h=0.72\pm 0.08$.
Column six tabulates the best fit parameters from combined bandpowers\protect\cite{wang}. 
Fits are to the CMB data alone with $95\%$ confidence limits.
\label{tab:table1}}
\begin{center}
\footnotesize
\begin{tabular}{|l|ccccc|}
\hline
{}&Toco\protect\cite{Miller2001}&BOOM\protect\cite{deBernardis_inst}&
MAXIMA\protect\cite{Lee}&DASI\protect\cite{Leitch}&Comb. \\
\hline
Platform&ground&balloon&balloon&ground&-- \\
Type&beam-switch&mapping&mapping&int&-- \\
Detector&HEMT,SIS&Bolo&Bolo&HEMT&-- \\
Freq (GHz)&{30-144}&90-400&150-410&26-36&-- \\
resolution&$0.2^{\rm o} - 1^{\rm o}$&$0.17^{\rm o} - 0.23^{\rm o}$&$10\prime$&$20\prime$&-- \\
Sky (deg)$^2$&500&1800&124&400&-- \\
\hline
$\Omega_{TOT}$&--&$1.03\pm0.06$&$0.90\pm0.15$&$1.04\pm0.06$&-- \\
$\Omega_{K}$&0&--&--&--&$-0.06^{+.13}_{-0.59}$ \\
$\Omega_m$&--&$0.50\pm0.20$&$0.25-0.5$&$0.4\pm0.15$&-- \\
$\Omega_{\Lambda}$&0.57&$0.52^{+0.24}_{-0.19}$&$0.4-0.76$&$0.6\pm0.15$&
	$0.60^{+0.32}_{-0.55}$ \\
$\Omega_{CDM}h^2$&0.16&$0.12\pm0.05$&$0.13\pm0.10$&$0.14\pm0.04$&$0.10^{+0.07}$ \\
$\Omega_{\rm b}h^2$&0.019&$0.021^{+0.004}_{-0.003}$&$0.025\pm0.01$&
	$0.022^{+0.004}_{-0.003}$&$0.02^{+0.06}_{-0.01}$ \\
$n_s$&1.18&$0.93^{+0.1}_{-0.08}$&$0.99\pm0.09$&$1.01^{+0.08}_{-0.06}$&$0.93^{+0.75}_{-0.16}$ \\
\hline
\end{tabular}
\end{center}
\end{table}

\section{The Future of the CMB}

While tremendous progress has been made using the CMB to measure cosmological
parameters, much
remains to be done. In terms of primary anisotropies the error bars and calibration
uncertainties over the range of measured angular scales need to be reduced, the detailed shape
of the damping tail at small angular scales has yet to be characterized, and the
range of angular scales between the COBE measurement and the more recent measurements
have yet to be connected by a single instrument. In addition, more work  
needs to be done in order to fully understand the extent to which foregrounds such
as point sources, dust, and synchrotron radiation are capable of contaminating CMB
results. Other foregrounds such as the Sunyaev-Zel'dovich effect (SZE) are capable of
providing powerful probes of the early universe in their own right. Furthermore,
the same processes responsible for creating the temperature anisotropies discussed throughout
this paper should also produce polarization anisotropies. The spectrum of these
anisotropies will provide a complementary look at the early universe and will allow
the removal of degeneracies inherent to measurement of the temperature anisotropy
spectrum alone. 

Experiments designed to measure primary anisotropies, polarization anisotropies, and the 
Sunyaev-Zel'dovich effect are underway. 
NASA's MAP satellite, launched on June 30th, 2001 will provide detailed information
about the spectrum of primary anisotropies out to $l\approx 1000$. It is also
expected to detect the polarization of the CMB over a wide range of multipoles. ESA's Planck
satellite, scheduled for launch in 2007 will measure the spectrum of both 
primary and polarization anisotropies up to $l\approx 2000$. Planck also
expects to survey the Sunyaev-Zel'dovich effect in clusters of galaxies. In the mean
time, several groups are analyzing data already obtained and many balloon-based and 
ground-based experiments are being designed to measure CMB polarization and to conduct non-targeted SZE surveys. Detector technology continues to improve
and new techniques are rapidly being developed. The near future
in this field promises to be as exciting as the recent past.

\section*{Acknowledgments}
The author would like to thank Erik Leitch for a thorough reading of the text, Mark
Devlin and Andreas Berlind for helpful comments, and is supported by Hubble Grant 
ASTR/HST-HF-0113.

\end{document}